\documentclass{aa} 
\usepackage{natbib}
\usepackage[dvips]{graphicx}
\begin{document}
   \title{3D photospheric velocity field of a Supergranular cell}
   \author{Del Moro, D., Giordano, S. and Berrilli, F.}
   \institute{Dipartimento di Fisica, Universit\`a di Roma ``Tor Vergata'', I-00133 Roma, Italy}
   \date{Received {\it date will be inserted by the editor}; accepted {\it date will be inserted by the editor} }
   \offprints{delmoro@roma2.infn.it}
   \authorrunning{Del Moro \emph{et al.}}
   \titlerunning{3D photospheric velocity field of a SG cell}
\abstract
{}
{We investigate the plasma flow properties inside a Supergranular (SG) cell, in particular its interaction with small scale magnetic field structures.}
{The SG cell has been identified using the magnetic network (CaII wing brightness) as proxy, applying the Two-Level Structure Tracking (TST) to high spatial, spectral and temporal resolution observations obtained by IBIS.
The full 3D velocity vector field for the SG has been reconstructed at two different photospheric heights.
In order to strengthen our findings, we also computed the mean radial flow of the SG by means of cork tracing.
We also studied the behaviour of the horizontal and Line of Sight plasma flow cospatial with cluster of bright CaII structures of magnetic origin to better understand the interaction between photospheric convection and small scale magnetic features.}
{The SG cell we investigated seems to be organized with an almost radial flow from its centre to the border.
The large scale divergence structure is probably created by a compact region of constant up-flow close to the cell centre.
On the edge of the SG, isolated regions of strong convergent flow are nearby or cospatial with extended clusters of bright CaII wing features forming the knots of the magnetic network.}
{}
\maketitle
\section{Introduction}
Solar research is currently working on understanding how turbulent convection on the Sun transports mass and energy through the convective zone, how it couples with the magnetic field and how it manages to deposit in the higher parts of the solar atmosphere the energy released from the corona.
Among the different approaches to these questions, observations of the solar photosphere are essential, as they provide the only direct look at what is happening just below the solar surface.
The hierarchy of surface features found on the photosphere are the visible representation of the plasma flows beneath the photosphere and are customarily classified by size and lifetime as patterns of granulation (1 Mm, 0.2 hr), mesogranulation (5-10 Mm, 5 hr) and supergranulation (15-35 Mm, 24 hr).
These features have been initially regarded as direct manifestation of various sized convection cells existing in the convection zone \citep{schrijver97,Raju97}; lately, the idea is consolidating that meso and supergranulation are signatures of a collective interaction of granular cells \citep{Rast03,Roudier03,berrilli05}.
Despite years of intensive studies, the character of their motions remains not completely understood \citep{beck00,krishan02,berrilli03,delmoro03,derosa04}.
The aim of the study we present is to investigate the origin of the supergranular (SG) flow field: directly convective or a collective interaction of smaller convective features.\\
The study performed by \citet{simon64} initiated the campaign to characterize supergranular flows.
Outflows on SG scales have been observed to sweep embedded granules and magnetic flux elements toward convergence lanes between cells \citep{leighton64,zwaan78,rimmele89,shine00}.
Such behaviour causes the chromospheric transition CaII k line to be a good proxy for the network of intercellular lanes due to the higher magnetic elements density in the SG perimeters.
The advent of full-disk Doppler imaging, provided by the MDI onboard SOHO spacecraft, has considerably improved our capability to study such features \citep{hathaway02,derosa04,paniveni04,meunier06}; but  direct observations of supergranular flows are still hindered by the fact that there is no contrast on supergranular scales in visible light, observations in CaII only provide the cell network boundaries and Doppler images show SG only away from disk centre.\\
At present, the only methods to reconstruct the full 3D vector velocity field are direct Doppler measurement in combination with a tracking type measure for the velocity horizontal component (above the $\tau=1$ surface) or Local Helioseismology (below the $\tau=1$ surface).
To gain complete insight of the dynamics of the plasma flows inside a SG structure, we need a spatial and temporal resolution still not reached by local helioseismology, while to obtain the 3D velocity field through the other method, observations with very high spatial, spectral and temporal resolution are necessary.
With the assumption that granule motions are mainly driven by plasma flows \citep{Rieutord01}, it is possible to employ the TST to infer the horizontal velocity field.\\
In this work we reconstruct the 3D velocity field of a single SG structure and investigate in detail its plasma flow using data acquired with the IBIS spectrometer, trying to discern whether the SG pattern has a convective nature or is originated by small scale structure interaction.
\section{Observations}
\begin{figure*}[!ht]
\begin{center}
\includegraphics[width=16.0cm, height=10.5cm]{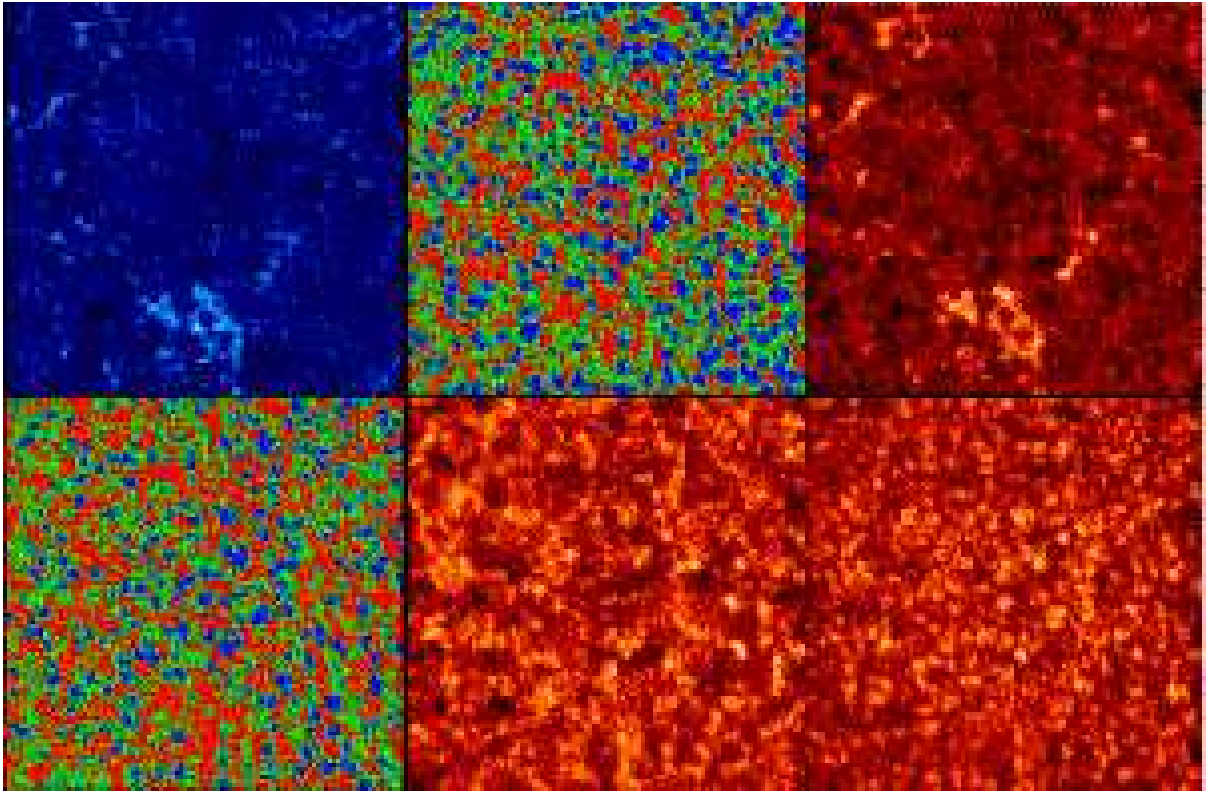}
\end{center}
\caption{
A representative synoptic panel from the $16^{th}$ October 2003 dataset.
Upper left panel: Ca II wing intensity image.
Upper middle panel: Doppler velocity field computed from FeI 709.0 nm line scan.
Upper right panel: FeI 709.0 nm line core intensity
Lower left panel: Doppler velocity field computed from FeII 722.4 nm line scan.
Lower middle panel: FeII 722.4 nm line core intensity.
Lower right panel: Continuum (near 709.0 nm) intensity image.
}
\label{synop}
\end{figure*}
The data utilized in this analysis have been acquired with the IBIS (Interferometric BIdimensional Spectrometer) 2D spectrometer \citep{Cavallini01, Cavallini06} on October 16, 2003 (from 14:24 UT to 17:32 UT).\\
We imaged a roundish network cell near the solar disk centre (SLAT=7.8°N, SLONG=3.6°E).
When observed in MDI high-resolution magnetograms, all the features outlining the cell exhibit negative polarity and seem to survive for at least 10 hours, with little or no evolution.\\
The full dataset consists of 600 sequences, containing a 16 image scan of the FeI 709.0 nm line, a 14 image scan of the FeII 722.4 nm line and 5 spectral images in the wing (line centre + 12 nm) of the CaII 854.2 nm line, imaging a round Field of View (FoV) of about 80" diameter.
Each monochromatic image was acquired with a 25 ms exposure time by a 12bit CCD detector, whose pixel scale was 0.17$\arcsec\cdot$pixel$^{-1}$.
The time required for the acquisition of a single sequence was 19 s, thus setting the temporal resolution.
Each image was reduced with the standard IBIS pipeline \citep{katja06, giordano07}, correcting for CCD non linearity effects, dark current, gain table and blue shift.
The Line of Sight (LoS) velocity fields were computed for the Fe I and Fe II lines by means of Doppler shifts, evaluated, pixel by pixel, fitting a Gaussian on the line profile.
In order to remove the orbital contribution, we set to zero the average value of each LoS velocity image.
The 5-minutes oscillations were removed applying a 3D Fourier filter in the $k_h-\omega$ domain with a cut-off velocity of 7 km$\cdot$s$^{-1}$ both on intensity and velocity image series.\\
After the whole reduction process and selecting only the period of good seeing we are left with a 30 minutes dataset imaging a square FoV of $\sim 50"$. An example of the images of this reduced dataset is shown in Fig. \ref{synop}.
The mean resolution due to the seeing of the CaII images is 0.35"; the mean resolution of the LoS velocity images, the intensity continuum images and the line core images is instead 0.45", somewhat degraded by both the reduction pipeline and the $k_h-\omega$ filtering.\\
In order to obtain information about the depth dependence of a photospheric quantity by associating a suitable `formation zone' with a line, it is possible to consider its effect on the line characteristic as linear perturbations and to study the Response Function \textit{RF} of the emergent line characteristic at the observed wavelengths within the line.
In particular, the \textit{RF}$_{p}^{I}$ is, at each depth, the function we must use to weigh the perturbation $p$ in order to get the variation of the emergent intensity $I$ \citep{Caccin77}.
\begin{figure}[!ht]
\includegraphics[width=8cm, height=7cm]{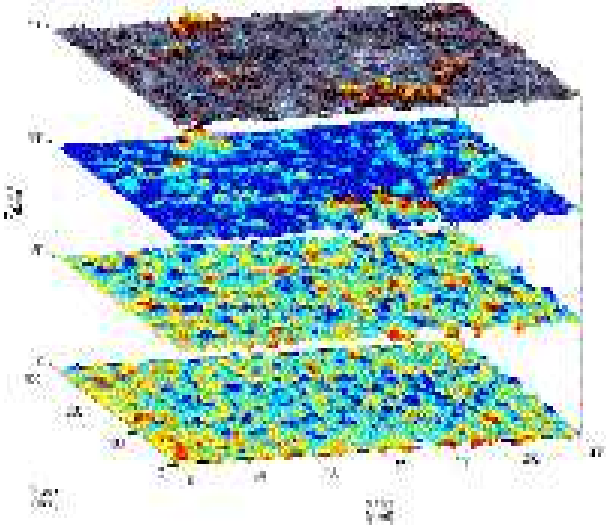}
\caption{Core intensity fields of FeII 722.4 nm (z$\simeq$50 km) and FeI 709.0 nm (z$\simeq$100 km ) in comparison with continuum image (z$\simeq$0 km) and CaII 854.2 nm wing intensity field (z$\simeq$150 km) \citep{CaII}.
The z axis is greatly exaggerated with respect to the x-y axes in order to allow a better visualization.}
\label{fig1}
\end{figure}
This approach has been employed to derive the \textit{RF}$^{I}_{T}$ and \textit{RF}$^{I}_{V}$ for the spectral lines FeI 709.0 nm and FeII 722.4 nm \citep{DelMoro05}.\\
In Table \ref{table1} we report the photospheric depths of the line core \textit{RF}$^{I}_{T}$ maximum ($z_{T_{core}}$), and of the mean \textit{RF}$^{I}_{V}$ maximum ($z_{V_{line}}$) for the two spectral lines.
We also report the \textit{RF} full width at half maximum  for the two spectral lines: these rather large values imply broad formation zones for both the FeI 709.0 nm and FeII 722.4 nm Doppler velocity and core intensity signals.
\begin{table*}[!ht]
\begin{center}
\begin{tabular}{lccccc}
\hline
Line &Wavelength &$z_{T_{core}}$ &FWHM$_{\textit{RF}^{I}_{T}}$ &$z_{V_{line}}$&FWHM$_{\textit{RF}^{I}_{V}}$ \\
 &[nm]&[km]&[km]&[km]&[km]\\
\hline
FeI &709.0 &$\simeq$100 &$\sim$300 &$\simeq$140 &$\sim$300\\
FeII &722.4 &$\simeq$50 &$\sim$200 &$\simeq$70 &$\sim$200\\
\hline
\end{tabular}
\end{center}
\caption{Line RF peak depths.
Depths are in km above the level $\tau_{500nm}=1$.}
\label{table1}
\end{table*}
\section{3D Velocity Field}
\begin{figure}[!ht]
\includegraphics[width=8cm, height=6cm]{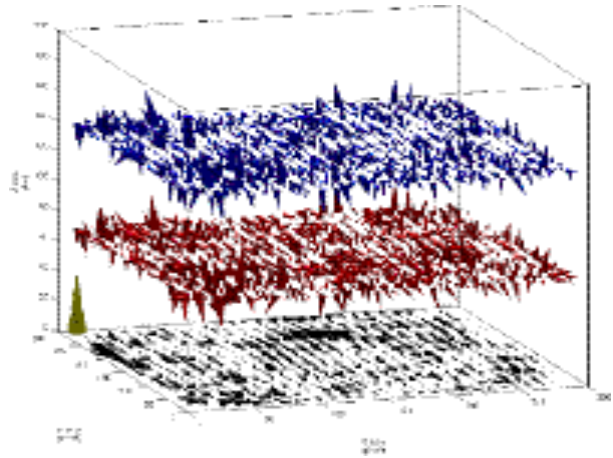}
\caption{3D representation of velocity vectors extracted from continnum (z$\simeq$0 km), FeII 722.4 nm (z$\simeq$70 km) and FeI 709.0 nm (z$\simeq$140 km).
Cone size is proportional to the velocity vector module: the yellowish cone corresponds to 1 km$\cdot$s$^{-1}$.
The z axis is greatly exaggerated with respect to the x-y axes in order to allow a better visualization.}
\label{fig2}
\end{figure}
\begin{figure}[ht!]
\begin{center}
\includegraphics[width=6cm]{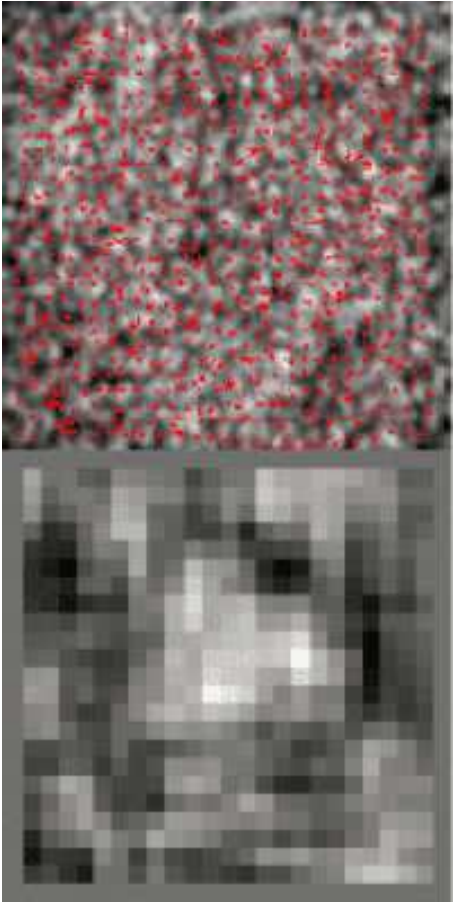}
\end{center}
\caption{
Upper panel: average continuum image with the horizontal velocity field (obtained by tracking granules) represented as red arrows.
The granules were tracked by applying the TST to the continuum image time series.
Lower panel: divergence field computed from the interpolated horizontal velocity field.
}
\label{I1}
\end{figure}

The TST procedure \citep{DelMoro04} has been applied on the continuum image series and on both the FeII 722.4 nm and FeI 709.0 nm Doppler field series, in order to retrieve the horizontal velocity field at different depths of the solar atmosphere.\\
To minimize the effect of the proper motion of the granules, which are used as trackers of the mean plasma flow, we computed the horizontal velocity field using all the structures that were tracked, so that statistically at least one tracker is present in each interpolated horizontal velocity field pixel, as suggested by \citet{behan00}.
This means we used a grid step of $\sim 1.5$ Mm and a temporal window of $\sim 30$ min.
Possibly, this would not completely remove the noise associated with granule proper motions or residuals from the 5-min oscillation filtering, but should minimize it.\\
Combining the horizontal velocity fields retrieved from the Dopplergrams by the TST and the Doppler LoS velocity, the 3D vector field has been reconstructed for the FeII 722.4 nm and FeI 709.0 nm lines.
We are aware that associating the vector fields to precise heights in the photosphere is an oversemplification, as can be readily understood from the large FWHM reported in Table \ref{table1}, nevertheless, we did it for the sake of a good visualization: in Fig. \ref{fig2} we show the mean 3D velocity field associated to the the dataset.\\
In both the 3D fields we retrieved, the vector velocity appears to be structured quite coherently with the SG feature visible in the CaII wing images.\\
In order to further investigate the structuring of the velocity field, we computed the average continuum (upper panel of Fig. \ref{I1}) and CaII wing (bottom panel of Fig. \ref{fig7}) intensity images, the average Dopplergram from FeII 722.4 nm (middle panel of Fig. \ref{fig7}) and the average Dopplergram from FeI 709.0 nm (upper panel of Fig. \ref{fig7}) and correlated them.
While the average continuum image does not show any evident signal, there is a significant correlation between strong downflows and bright CaII features, in particular for the complex cluster of features in the lower part of the FoV.
This issue can be at least partially explained by the coherence of the 3D velocity field with the SG structure.
We expanded this study by comparing the averaged images with the horizontal velocity field extracted by the TST from granules as seen in the continuum and up-flows from the FeI 709.0 nm Doppler images.\\
We excluded from this analysis the FeII 722.4 nm Doppler images because we found its horizontal velocity field to be not as reliable as the others.
This is due to the TST finding less than optimal number of features to track because of the shallowness of the FeII 722.4 nm line.
A shallow line Doppler shift is much harder to measure by the LoS velocity reconstruction procedure, resulting in a more noisy dopplergram.
This noise is interpreted by the TST as a fast variation of the structures, therefore causing a lot of them to be rejected for the tracking.
As a consequence, the TST finds too few trackers in the 722.4 Doppler for the divergence field to be reliably reconstructed.\\
\begin{figure}[ht!]
\begin{center}
\includegraphics[width=6cm]{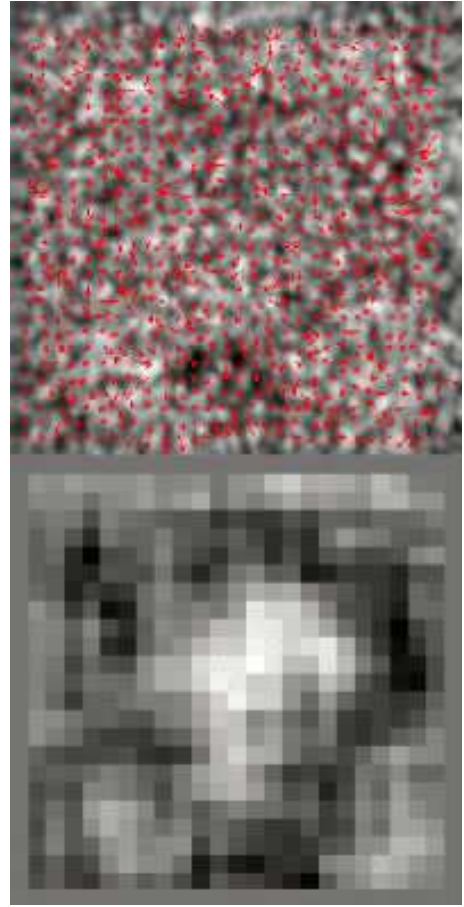}
\end{center}
\caption{
Upper panel: average FeI 709.0 nm Doppler velocity image with the horizontal velocity field (obtained by tracking up-flows) represented as red arrows.
The up-flows were tracked by applying the TST to the FeI 709.0 nm Doppler velocity field time series.
Lower panel: divergence field computed from the interpolated horizontal velocity field.
}
\label{V2}
\end{figure}
The horizontal velocity fields extracted by the TST are shown superimposed on the average images in the left panels of Fig. \ref{I1} and Fig. \ref{V2}, above the associated 2D divergence images. The values of the divergence fields range from $+0.25~km~s^{-1}~Mm^{-1}$ in the brightest part of the image to $-0.25~km~s^{-1}~Mm^{-1}$ in the darkest parts.
The continuum granules and FeI 709.0 nm up-flow fields show a divergent flow from the centre of the SG structure and convergent flows in the border of the SG structure.
In detail, these two fields agree very well, showing a single, large divergent feature in the centre of the SG, whose mean value is about $+0.1~km~s^{-1}~Mm^{-1}$, almost completely surrounded by convergent flows of the same magnitude.
The peak divergence signals we retrieved both in the centre and in the periphery of the SG cell are an order of magnitude larger than the averaged values reported by \citet{meunier06}.
This discrepancy probably stems mainly from the different temporal and spatial averaging processes in the divergence reconstruction and marginally from the different resolution of the two datasets.\\
The structuring of the divergence field is very compatible with a net flow from the centre of the SG to its border.\\
Moreover, examining the LoS velocity fields, we found a strong and stable up-flow region nearby the divergence maximum, with a mean FeII 722.4 nm Doppler velocity value of $V_c$ $\sim$200 m$\cdot$s$^{-1}$ for almost the whole time span.
This last region is liable to be the origin of the divergence signal we measured, possibly as suggested by \citet{Rieutord00, Roudier03}.\\
Observing the divergence images (bottom panels of Fig. \ref{I1} and Fig. \ref{V2}) extracted from the horizontal velocity fields, the supergranule is outlined by convergences on $\sim~66\%$ of its circumference, while the bright cluster area clearly visible in the CaII wing image does not seem to be a region of strong convergence, despite the fact that it is mostly formed of down-flows.\\
This region has a mean FeII 722.4 nm Doppler velocity value $V_c$ $\sim$-100 m$\cdot$s$^{-1}$ for the whole dataset time span (T $\simeq$ 0.5 hour), and it seems to mantain similar values also for the part of the observations discarded for the loss of spatial resolution due to worsening seeing condition.
A similar cluster of bright CaII structures is present in the upper-left part of the FoV, but its associated downflow shows a much smaller coherence: it has a mean FeII 722.4 nm Doppler velocity value $V_c$ $\sim$-50 m$\cdot$s$^{-1}$ for more than half of the time span, then it drops to $\sim$-20 m$\cdot$s$^{-1}$.
Whether or not downflow regions like these may be organizing the SG pattern, as predicted by \citet{Rast03}, is a question we cannot address due to the short duration of our dataset.\\
\section{Horizontal Flow Analysis via Cork Tracking}
To further extract information about the plasma motion inside the SG structure, we tracked the evolution of tracers (corks) passively advected by instantaneous velocity and intensity fields.
The corks, initially randomly spread over the FoV, are moved following the local gradient towards sites of minimum intensity or of minimum velocity in the case of intensity or velocity fields, respectively.
We will compare the final and initial positions of the corks, which will give us information about the motion of downflows in the field of view.
Corks are tracked for $\sim 16$ minutes (a time sufficiently longer than the characteristic time scale of photospheric fields \citep{DAN01,berrilli02} to let the cork settle in a downflow feature and track it for a while) and their initial and final position are stored.
As corks tend to accumulate in long lasting downflow structures, new corks are added each $\sim 5$ minutes in order to also track structures forming during the observations.
\begin{figure}[!ht]
\includegraphics[width=8cm, height=5cm]{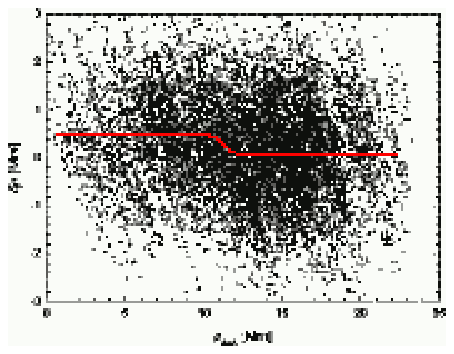}
\includegraphics[width=8cm, height=5cm]{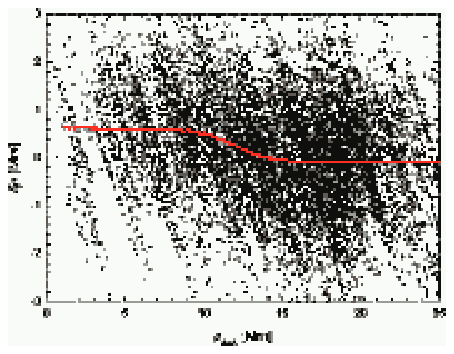}
\includegraphics[width=8cm, height=5cm]{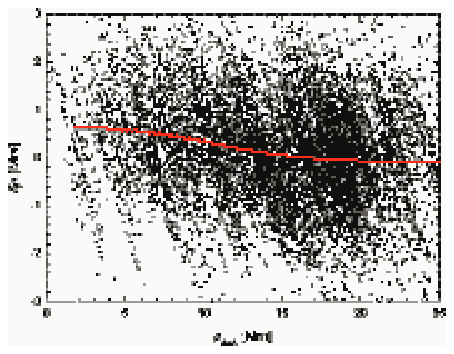}
\caption{Cork displacements versus initial positions.
Top Panel: results of the cork tracking for the intensity field from continuum images.
Central Panel: results of the cork tracking for the vertical velocity field extracted from FeII 722.4 nm.
Bottom Panel: results of the cork tracking for the vertical velocity field extracted from FeI 709.0 nm.
Each scatter plot has been fitted with a sigmoidal function (equation \ref{eq1}). The retrieved fits are overplotted on the relative scatter plots.}
\label{cork_g}
\end{figure}
In Fig. \ref{cork_g} we report the result of the cork tracing for a continuum image series and for both the FeII 722.4 nm and the FeI 709.0 nm Doppler fields.
In particular, we plotted the difference between the final and initial distances from the image centre of the corks versus their initial distances from the image centre.
The alignment effect of the scatter plot is due to an inverse linear relationship between the $\rho_{start}$ and the $\delta\rho$ of corks with different initial postions which end in the same `attractor' and therefore share their final position.\\
As the image is centred on the SG structure, this will give us information about a possible difference of mean flows inside and outside the SG.
In order to investigate the properties of the distribution, we fit on the scatter plots a sigmoidal function:
\begin{equation}
				y=\frac{A_1-A_2}{1+ e^{(x-x_0)/dx}} +A_2
\label{eq1}
\end{equation} 
so that $x_0$ will tell where the transition between the two values $A_1$ and $A_2$ of the distribution takes place and $dx$ will tell how fast this transition is.
The parameters of the fit are retrieved by a recursive Levenberg-Marquardt minimization algorithm.
\begin{table*}[!ht]
\begin{center}
\begin{tabular}{lccccc}
\hline
   &$A_1$ &$A_2$ &$x_0$&$dx$\\
   &[Mm]&[Mm]&[Mm]&[Mm] \\
\hline
Continuum Intensity&$0.49 \pm 0.02$&$0.08 \pm 0.01$&$11.3 \pm 0.2$&$0.3 \pm 0.2$\\
FeII 722.4 LoS Velocity &$0.61 \pm 0.03$&$-0.09 \pm 0.02$&$12.0 \pm 0.2$&$1.1 \pm 0.2$\\
FeI 709.0 LoS Velocity &$0.67 \pm 0.08$&$-0.09 \pm 0.03$&$10.9 \pm 0.7$&$2.9 \pm 0.6$\\
\hline
\end{tabular}
\end{center}
\caption{Parameters of the sigmoidal fits to the scatter plots reported in Fig. \ref{cork_g}.
}
\label{table2}
\end{table*}
The fits agree in retrieving positive values of $A_1$ and near zero values for $A_2$ (Table \ref{table2}).
This means that the corks inside a circle of radius $x_0$ from the image centre tend to increase their radial distance, while the corks outside have no preferred direction in their motion.
Several simulations on randomly generated velocity fields showed that we can neglect the contribute from corks whose initial position is so near to the image centre that they are biased towards positive radial displacement.
Finally, we tested the robustness of these results against the initial guesses and against the SG centre position in the FoV. The retrieved parameters do not depend on these factors, as long as the initial guesses are of the same order of magnitude of the convergence values or the FoV shift is less than 2.5 Mm.\\
\begin{figure}[!ht]
\begin{center}
\includegraphics[width=7cm, height=7cm]{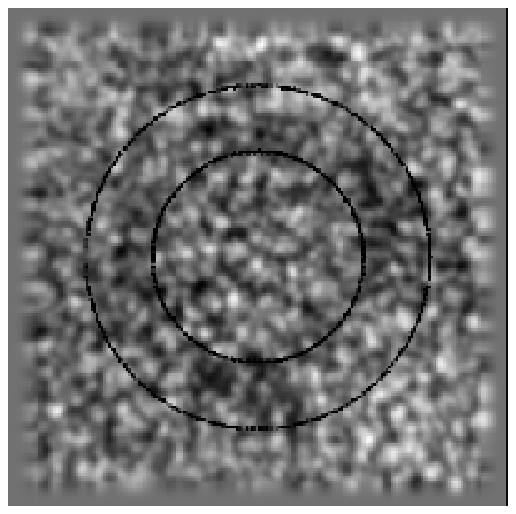}
\includegraphics[width=7cm, height=7cm]{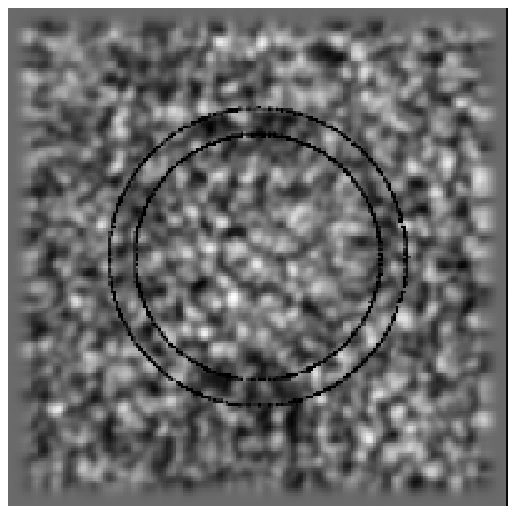}
\includegraphics[width=7cm, height=7cm]{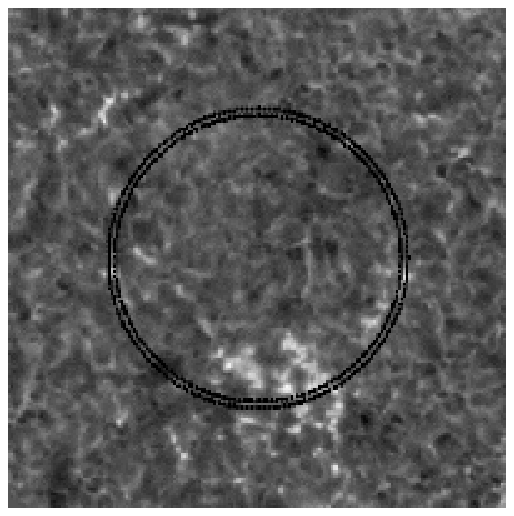}
\caption{Mean images with superimposed the SG dimension extracted from the cork tracking.
Top Panel: mean FeI 709.0 nm Doppler image with SG extracted from from  FeI 709.0 nm Dopplergrams ($\sim 140$ km).
Central Panel: mean FeII 722.4 nm Doppler image with SG extracted from from FeII 722.4 nm Dopplergrams ($\sim 70$ km).
Bottom Panel: mean CaII 854.2 nm wing image with SG extracted from the intensity continuum images ($\sim 0$ km).}
\end{center}
\label{fig7}
\end{figure}
In Fig. \ref{fig7} we show the mean intensity fields associated to the plots in Fig. \ref{cork_g}, with superimposed the location of the change of the $A$ value represented as an annulus of mean radius $x_0$ and thickness $2dx$.
The three annular shapes essentially agree in retrieving the same SG diameter of $\sim~25~Mm$.\\
The width of the annuli, instead, seems to depend on the atmospheric altitude.
In the upper panel of Fig. \ref{cork_p} we report the value of $dx$ computed by the sigmoidal fits versus the photospheric height.
Error bars represent the standard deviation from the fit.\\
Recently, \citet{berrilli02} found a similar height dependence of the statistical properties of granular flows.
In particular, they reported an intense braking in the first $\sim 120$ km of the photosphere, confirmed by \citet{puschmann06} and a damping effect that filtered out small features in higher atmospheric layers, letting only large flow features penetrate into the upper photosphere.\\
The same process can explain the broadening of the SG border we found: in higher layers the corks are collected in larger and fewer downflow structures.
As more corks are collected by the same structures, the number of independent tracers is decreased and similarly the precision of the retrieval of the boundary is decreased.\\
Instead, we can exclude that such a smoothing effect is due to data reduction or seeing, because in that case the SG border retrieved from the FeI 709.0 nm LoS field would have been thinner than the one retrieved from the FeII 722.4 nm LoS field, as the latter shows lower contrast features, more prone to be degraded by the loss of spatial resolution.\\
\begin{figure}[!hb]
\includegraphics[width=8cm, height=5cm]{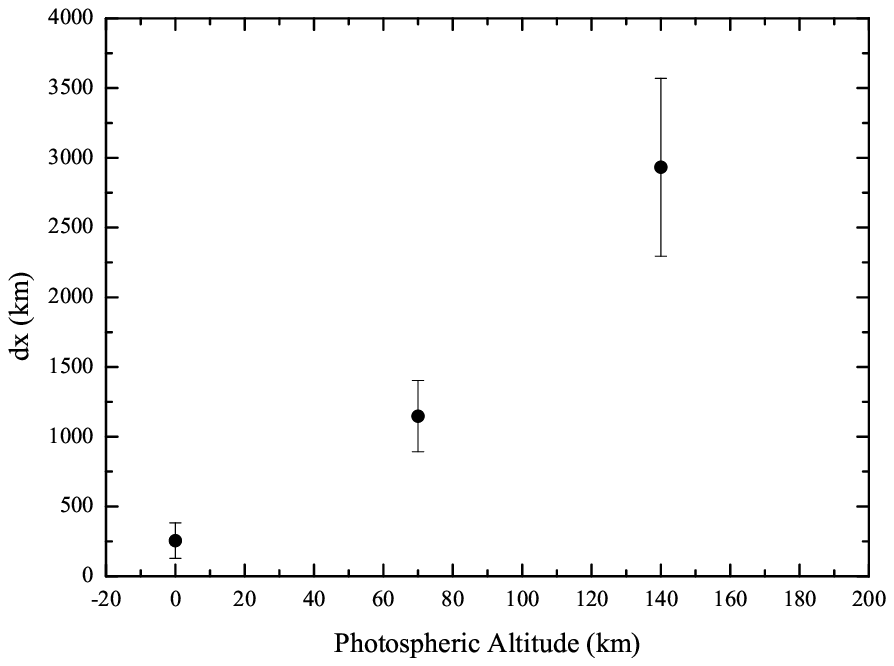}
\includegraphics[width=8cm, height=5cm]{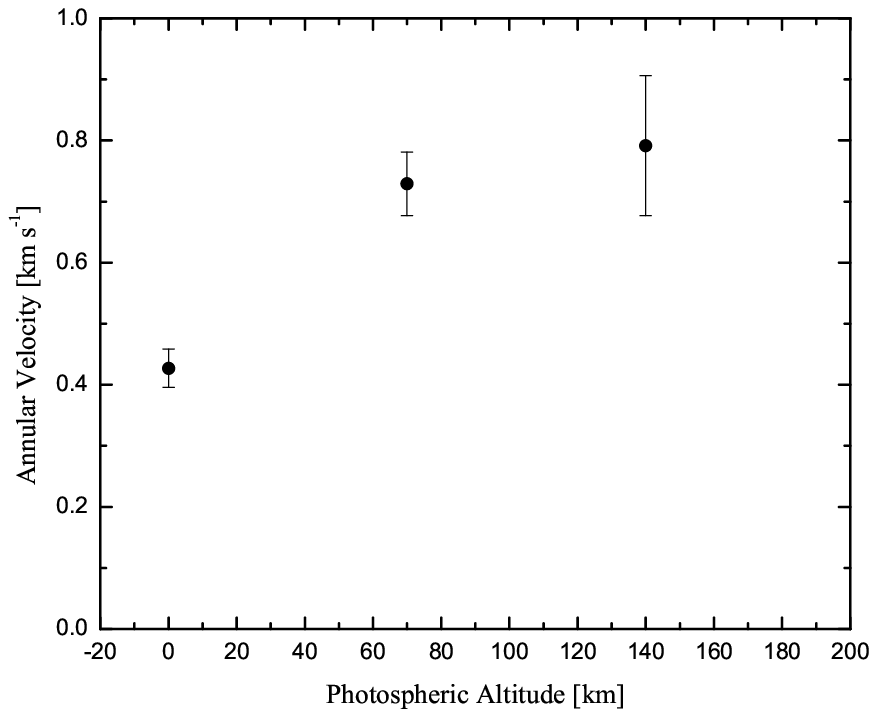}
\caption{Top panel: $dx$ (annulus width) versus photospheric height.
Bottom panel: radial velocity versus photospheric height.
}
\label{cork_p}
\end{figure}
Due to the form of equation \ref{eq1}, the difference $A_1~-~A_2$, divided for the time allotted to the corks to move, will give the mean radial velocity experienced by the corks.
We plot in the bottom panel of Fig. \ref{cork_p} the radial velocity retrieved from the three scatter plots as a function of the photospheric altitude. Error bars represent the standard deviation from the fit.\\
To account for these results, we assume that the large and more coherent features present in the LoS dopplergrams are reliable to retrieve the radial velocity measure, while the measure from the continuum images is somewhat reduced by the presence of tiny structures which are more turbulent in their motion.
Such structures are not present in the higher layers dopplergrams because of the damping effect already discussed.\\
We therefore discard the value obtained from the WL dataset because it is probably smeared by the turbulent motions of very small scale features and take into account only the two values retrieved from the higher layers, retrieving a mean velocity of $0.75\pm~0.05~km~s^{-1}$.\\
Such a value for the flows from the SG structure centre is consistent with the literature \citep{simon64,hathaway02,paniveni04,meunier06}.\\
\section{Conclusions}
The study of the full 3D velocity field of a SG shows that strong downflows are located on the border of the supergranular structure, but also that the mean granular flow regresses from the centre to the periphery of the SG.\\
The divergence images show that the SG structure is outlined by convergence sites on $\sim~66\%$ of its border.
The retrieved divergence values show a nearly radial flow of $\sim0.1~km~s^{-1}~Mm^{-1}$ from the centre of the SG and convergent flows of the same magnitude in its border.\\
The analysis of the evolution of passive tracers on intensity and velocity fields shows that inside the SG structure there is a preferential radial flow towards the SG border of $0.75\pm~0.05~km~s^{-1}$.\\
The height behaviour of the thickness of the SG border, again retrieved via cork tracing, shows an increase of the border width with height.
This is probably due to a filtering effect with height, which preferentially allows large flow features to penetrate into the upper photospheric layers.\\
The large and CaII bright cluster of structures in the lower part of the FoV, is not a site of strong convergence, but is a site of long-lasting downflows.
We also found a strong and stable upflow nearby the centre of the cell, liable to organize the CaII bright structures by sweeping them out of the SG cell.\\
The result presented in this paper are extracted from a subset of a longer timeseries of excellent spectral and temporal resolution, but varying spatial quality due to seeing.
The used 30 min subset is characterized by a constant and good spatial resolution.
This allowed us to detect precisely the flow associated with the SG.
Anyhow, our analysis would have greatly benefited from a longer time sequence and other SG structures to analyze. In the future, we plan to apply this analysis to a collection of SG structures, so as to derive some statistical describer and possibly generalize the results.
\begin{acknowledgements}
We thank the referee, T. Roudier, for suggestions and comments that have signiﬁcantly improved this paper.
Part of this work was supported by Rome ``Tor Vergata'' University Physics Department grants. The data were acquired by instruments operated by the National Solar Observatory.
The National Solar Observatory is a Division of the National Optical Astronomy Observatories, which is operated by the Association of Universities for Research in Astronomy, Inc., under cooperative agreement with the National Science Foundation.
DDM thanks the High Altitude Observatory for support and C. Sormani for helpful comments.
The authors aknowledge k. Janssen for the development of the IBIS data reduction pipeline, V. Penza for the calculation of the line RFs and M. Rast for very useful discussions and comments.
\end{acknowledgements}


\begin{thebibliography}{}
\bibitem[Behan(2000)]{behan00} Behan, A.
	2000, Proceedings of the 19th ISPRS Congress and Exhibition - Geoinformation for All. Amsterdam, The Netherlands, 16th - 23rd July 2000.
\bibitem[Beck \& Duvall(2000)]{beck00} Beck, J. G., Duvall, T. L., Jr.
	2000, BAAS, 32, 802
\bibitem[Berrilli \emph{et al.}(2002)]{berrilli02} Berrilli, F., Consolini, G., Pietropaolo, E., Caccin, B., Penza, V., Lepreti, F.
	2002, \aap, 381, 253
\bibitem[Berrilli \emph{et al.}(2004)]{berrilli03} Berrilli, F., Del Moro, D., Consolini, G., Pietropaolo, E., Duvall, T. L., Jr., Kosovichev, A. G.
	2004, Sol. Phys., 221, 33
\bibitem[Berrilli \emph{et al.}(2005)]{berrilli05} Berrilli, F., Del Moro, D., Russo, S., Consolini, G., Straus, Th.
	2005, \apj, 632, 677
\bibitem[Caccin \emph{et al.}(1977)]{Caccin77} Caccin, B., Gomez, M.~T., Marmolino, C. \& Severino, G.,
	1977, \aap, 54, 227
\bibitem[Cavallini \emph{et al.}(2001)]{Cavallini01} Cavallini, F., Berrilli, F., Cantarano, S., Egidi, A.
	2001, Mem. SaIt, 72, 554
\bibitem[Cavallini(2006)]{Cavallini06} Cavallini, F.
	2006, Sol. Phys., 236, 415
\bibitem[Del Moro \emph{et al.}(2004)]{delmoro03} Del Moro, D., Berrilli, F., Duvall, T. L., Jr., Kosovichev, A. G.
	2004, Sol. Phys., 221, 23
\bibitem[Del Moro(2004)]{DelMoro04} Del Moro, D.
	2004, \aap, 428, 1007
\bibitem[Del Moro(2005)]{DelMoro05} Del Moro, D.
	2005, PhD Thesis
\bibitem[DeRosa \& Toomre(2004)]{derosa04} DeRosa, M. L., Toomre, J.
	2004, \apj, 616, 1242
\bibitem[Deubner(1971)]{deubner71} Deubner, F.-L.
	1971, Sol. Phys., 17, 6
\bibitem[Frazier(1970)]{frazier70} Frazier, E.N.
	1970, Sol. Phys., 14, 89
\bibitem[Giordano \emph{et al.}(2007)]{giordano07} Giordano, S. Del Moro, D. Berrilli, F.
	2007, submitted
\bibitem[Hathaway \emph{et al.}(2002)]{hathaway02} Hathaway, D. H., Beck, J. G., Han, S., Raymond, J.
	2002, Sol. Phys., 205, 25
\bibitem[Janssen \& Cauzzi(2006)]{katja06} Janssen, K., Cauzzi, G.
	2006, \aap, 450, 365
\bibitem[Krishan \emph{et al.}(2002)]{krishan02} Krishan, V., Paniveni, U., Singh, Jagdev, Srikanth, R.
	2002, MNRAS, 334, 230
\bibitem[Leighton(1964)]{leighton64} Leighton, R. B.
	1964, \apj, 140, 1547
\bibitem[Meunier \emph{et al.}(2007)]{meunier06} Meunier, N., Tkaczuk, R., Roudier, Th., Rieutord, M.
	2007, \aap, 461, 1141
\bibitem[M\"uller \emph{et al.}(2001)]{DAN01} M\"uller, D.A.N., Steiner, O., Schlichenmaier, R., Brandt, P.N.
        2001, Sol. Phys., 203, 211
\bibitem[Musman \& Rust(1970)]{musman70} Musman, S., Rust, D.S.
	1970, Sol. Phys., 13, 261
\bibitem[Paniveni \emph{et al.}(2004)]{paniveni04} Paniveni, U., Krishan, V., Singh, Jagdev, Srikanth, R.
	2004, MNRAS, 347, 1279
\bibitem[Puschmann \emph{et al.}(2005)]{puschmann06} Puschmann, K. G., Ruiz Cobo, B., Vázquez, M., Bonet, J. A., Hanslmeier, A.
	2005, \aap, 441, 1157
\bibitem[Qu \& Xu (2002)]{CaII} Qu, Z. \& Xu, Z.,

   	2002, Chin. J. Astron. Astrophys., 2, 71
\bibitem[Rast(2003)]{Rast03} Rast, M. P., 
	2003, \apj, 597, 1200
\bibitem[Raju \emph{et al.}(1999)]{Raju97} Raju, K. P., Srikanth, R., Singh, Jagdev
	1999, BASI, 27, 65
\bibitem[Rieutord \emph{et al.}(2000)]{Rieutord00} Rieutord, M., Roudier, Th., Malherbe, J.M., Rincon, F.
	2000, \aap, 357, 1063
\bibitem[Rieutord \emph{et al.}(2001)]{Rieutord01}  Rieutord, M., Roudier, Th., Ludwig, H. G., Nordlund, \AA{}., Stein, R.
	2001, \aap, 377, L14
\bibitem[Rimmele(1989)]{rimmele89} Rimmele, T., Schroeter, E. H.
	1989, \aap, 221, 137
\bibitem[Roudier \emph{et al.}(2003)]{Roudier03} Roudier, Th., Lignieres, F., Rieutord, M., Brandt, P.N., Malherbe, J.M.
	2003, \aap, 409, 301
\bibitem[Schrijver \emph{et al.}(1997)]{schrijver97} Schrijver, C. J., Hagenaar, H. J., Title, A. M.
	1997, \apj, 475, 328 
\bibitem[Shine \emph{et al.}(2000)]{shine00} Shine, R. A., Simon, G. W., Hurlburt, N. E.
	2000, Soph, 193, 313
\bibitem[Simon \& Leighton(1964)]{simon64} Simon, G. W., Leighton, R. B.
	1964, \apj, 140, 1120
\bibitem[Zwaan(1978)]{zwaan78} Zwaan, C.
	1978, Sol. Phys., 60, 213
\end{thebibliography}
\end{document}